%% file: Main.tex
\documentclass[10pt,journal,compsoc]{IEEEtran}
\usepackage{ifpdf}
\usepackage[nocompress]{cite} %
\usepackage[pdftex]{graphicx}
\usepackage{amsmath}
\usepackage{amsfonts}
\usepackage{amssymb}
\usepackage{bm}
\usepackage{upgreek} %
\usepackage{enumitem}
\usepackage[ruled,vlined,linesnumbered]{algorithm2e}
\SetKwInput{KwInput}{Input}
\SetKwInput{KwOutput}{Output}
\SetAlgoNlRelativeSize{0}
\usepackage{array}
\usepackage[caption=false,font=footnotesize,labelfont=sf,textfont=sf]{subfig}
\usepackage{stfloats}
\usepackage{balance}
\usepackage{url}
\usepackage{xcolor}
\usepackage{xspace}
\usepackage[normalem]{ulem}
\begin{document}
\bstctlcite{IEEEBSTcontrol}
\title{\huge Retrofitting Multi-Agent Systems for Harsh Communication Environments via Latent Dynamics Prediction}

\author{Jingzhi~Hu,~\IEEEmembership{Member,~IEEE,}
Xinyuanmeng~Yao,
Yuntao~Wang,~\IEEEmembership{Member,~IEEE,}
Yiliang~Liu,
Zhou~Su,~\IEEEmembership{Senior~Member,~IEEE,}
and~Jun~Luo,~\IEEEmembership{Fellow,~IEEE}%
\IEEEcompsocitemizethanks{%
\IEEEcompsocthanksitem J. Hu, Y. Wang, Y. Liu, and Z. Su are with the School of Cyber Science and Engineering, Xi'an Jiaotong University, Xi'an, China (e-mail: \{jingzhi.hu, yuntao.wang, liuyiliang\}@xjtu.edu.cn, zhousu@ieee.org).
\IEEEcompsocthanksitem X. Yao is with the School of Cyber Science and Engineering, Ningbo University of Technology, Ningbo, China (e-mail: yaoxym@nbut.edu.cn).
\IEEEcompsocthanksitem J. Luo is with the College of Computing and Data Science, Nanyang Technological University, Singapore (e-mail: junluo@ntu.edu.sg).}}

\IEEEtitleabstractindextext{\input{sections/1_abstract}}

\IEEEpubid{\parbox{\textwidth}{\centering\normalfont\footnotesize This work has been submitted to the IEEE for possible publication. Copyright may be transferred without notice, after which this version may no longer be accessible.}}
\maketitle
\IEEEdisplaynontitleabstractindextext
\IEEEpeerreviewmaketitle

\input{sections/2_introduction}
\input{sections/3_preliminaries}

\input{sections/4_HERALD}
\input{sections/5_experiment_setup}
\input{sections/6_evaluation}
\input{sections/7_conclusion}
\balance %
\input{Main.bbl}

\end{document}

%% file: sections/1_abstract.tex
\begin{abstract}
Mobile multi-agent systems (MAS) overcome the partial observability of individual agents by communicating their local states through data packets over wireless links. 
Under severe interference and mobility-induced intermittent connectivity in harsh environments, communication may suffer extreme packet loss or even adversarial manipulation, cutting agents off from timely teammate information, misleading their decisions, and breaking down collaboration.
We propose \emph{Herald}, a lightweight neural module deployed at each agent, which retrofits a MAS pretrained under ideal communication with resilience to harsh environments through latent dynamics prediction.
To reconstruct lost packets, Herald leverages the sequence-modeling capability of the Mamba-3 state-space model to capture the latent dynamics behind each agent's state, thereby predicting how it evolves across long and irregular gaps.
As for each packet received from another agent, Herald checks the state information it carries against the learned dynamics and the packets received from other teammates, screening out manipulated ones before they distort the tracker.
Moreover, Herald incurs no additional communication overhead and requires no parameter change to the underlying MAS.
Extensive experiments on the StarCraft multi-agent challenge demonstrate that Herald sustains 92.1\% of the ideal cooperative performance under a severe packet-loss rate of 90\%, and remains robust to multiple types of adversarial message manipulation.
\end{abstract}


%% file: sections/2_introduction.tex
\IEEEraisesectionheading{\section{Introduction}\label{sec:introduction}}
\IEEEPARstart{C}{ooperative} mobile multi-agent systems (MAS) are emerging as a key foundation for large-scale intelligent autonomy, with multi-agent reinforcement learning (MARL) providing a framework for learning their cooperative policies.
By integrating distributed perception, reasoning and control, multiple agents can accomplish tasks that are difficult or impossible for a single agent~\cite{Guo2024LargeLM}.
Representative applications include warehouse robot fleets~\cite{Li2021LifelongMA}, autonomous delivery robots~\cite{Srinivas2022AutonomousRD}, and UAV-based search-and-rescue teams.
Given that each mobile agent has access only to local and partial observations, communication is essential for sharing complementary information and coordinating decentralized decision-making.
This gives rise to \emph{communicative} MAS, in which each agent encodes its local observation history into a compact latent message, broadcasts it to teammates, and conditions its own decision on the messages received in return.

Nevertheless, in challenging environments such as underwater and underground deployments, communication links suffer from severe attenuation and interference, and become highly unreliable~\cite{Yan2025UnderwaterCoDesign,Pal2024WirelessUnderground}.
Agent mobility further aggravates the situation: as agents move, the distances, obstructions, and topology among them keep changing, causing fast-varying fading, blockage, and intermittent connectivity.
Extreme packet loss may deprive agents of timely teammate information, disrupting coordination and even causing collaboration to fail.
This problem is particularly acute in real-time decision-making, where communication is coupled to time-critical action selection and retransmission is often infeasible within the limited bandwidth and latency budget.
For example, in~\cite{Simoes2019EmergentCommunication}, the authors proposed asynchronous advantage actor-critic with communication (A3C2) and evaluated its performance under message loss, Gaussian noise, and message jumbling.
Such evaluations highlight the sensitivity of MAS to communication impairments.\looseness=-1\relax

Beyond packet loss, the open wireless medium further exposes this communication to adversaries: an adversary may intercept in-flight messages and replace them with manipulated ones~\cite{Tu2021AdversarialCommunication,Sun2023CertifiablyRobust}, or hijack teammates to broadcast forged ones under their legitimate identities~\cite{Li2024ByzantineRobust}.
Unlike a lost message that merely withholds information, a manipulated one actively misleads the receiver's decision and, once accepted, distorts its communication history and hence its interpretation of future messages.
Worse still, latent messages are opaque learned embeddings consumed directly by the policy, leaving the receiver little basis on which to distinguish manipulated ones from authentic ones.

Existing approaches to {communication impairments} in MAS can be broadly grouped into four categories.
\emph{Buffering and stale-message reuse} methods retain the latest successfully received message when a fresh packet is unavailable.
Temporal message control, for example, transmits only sufficiently novel messages and reuses buffered ones during packet loss~\cite{Zhang2020TMC}.
Although effective over short outages, this strategy becomes less reliable as the buffered information grows stale.
A separate line of work addresses \emph{delay compensation} by learning when to wait for delayed messages, weighting information according to its timeliness, or estimating delay-free information from communication histories~\cite{Yuan2023DACOM,Song2025CoDe,Fu2025RDC}.
These methods primarily target latency and outdated information rather than prolonged and irregular packet-loss gaps.
On the other hand, \emph{loss-aware robust training} methods inject communication impairments or lossy-link priors during MARL, enabling policies to rely more on reliable messages and suppress impaired ones~\cite{Yang2025CommunicationPriors}.
\emph{Missing-information prediction} methods recover unavailable peer observations, local states, or global representations through temporal, spatiotemporal, or model-based prediction~\cite{Santos2025MARO,Ding2026MATCH,Shi2026RMIOv2}.
These methods generally require the agent policy and the information predictor to be trained under the anticipated channel conditions.

All four categories implicitly presuppose that every successfully delivered message is authentic, and hence offer no defense once adversaries manipulate the delivered content.
Adversarial communication in MAS has been studied in isolation: attacks on the exchanged messages have been demonstrated~\cite{Tu2021AdversarialCommunication}, and existing defenses either train policies certifiably robust to such perturbations~\cite{Sun2023CertifiablyRobust} or build robustness to Byzantine teammates into the MARL formulation~\cite{Li2024ByzantineRobust}.
However, these defenses are each tied to the threat model assumed during training, so that a new adversarial environment calls for retraining the MAS from scratch. 
The challenge is further aggravated under extreme packet loss, where authentic messages are usually missing and the receiver is left with little evidence to identify manipulated ones, making a stable attack-resistant policy even harder to learn.\looseness=-1\relax

Moreover, the conventional approach to train a MAS to be robust to communication impairments is often prohibitively costly.
In the first place, MARL training is highly sensitive to the experimental setup, reward assignment, and hyper-parameters.
This makes identifying a suitable configuration and successfully obtaining an effective cooperative policy for MAS expensive in itself. 
Even with the recent advancement of world-model-based MARL~\cite{Liu2024MAZero,Zhang2025MARIE}, such cost is still pronounced: although imagined exploration substantially reduces the real sample requirement, jointly optimizing the world model and the action policies is more delicate and more prone to collapse~\cite{Zhang2025MARIE,Hansen2024TDMPC2,Hafner2025DreamerV3}.
Introducing extreme packet-loss conditions and adversarial attacks into this learning loop would further perturb the inputs and rollout distribution of the learned world model, and may therefore aggravate these optimization difficulties.
Consequently, an alternative research direction becomes more attractive and of practical importance, namely how to efficiently retrofit a MAS trained under an ideal communication environment with resilience to severe communication environments.\looseness=-1\relax

To address this, we propose \emph{Herald}, a lightweight neural module that retrofits \textbf{h}arsh communication \textbf{e}nvironment \textbf{r}esilience into \textbf{a}gents through \textbf{l}atent \textbf{d}ynamics prediction.
Herald exploits the fact that latent states in an MAS are constrained by its implicit latent dynamics, which is shaped by both the shared global environment and the common cooperative policy.
To capture the latent dynamics, Herald builds a latent-dynamics tracker based on the Mamba-3 architecture~\cite{lahoti2026mamba3}, which can capture the latent dynamics in a computationally efficient manner while remaining accurate over the long gaps caused by extreme packet loss.
Guided by the captured latent dynamics, we develop a predictor, which combines the latent states of the agent itself and those received from its teammates to predict any lost messages. 
Along with each prediction, the predictor also outputs a confidence estimate that regulates how much the prediction can be relied upon. 
Moreover, leveraging the same latent dynamics, we design a verifier, which first assesses individually the trustworthiness of each received message, i.e., the likelihood that it is authentic.
The verifier further refines the assessment by jointly considering all received messages with awareness of their individual trustworthiness, enabling Herald to effectively detect and reject adversarial messages.

We develop a two-stage distillation algorithm to train Herald efficiently.
Stage 1 is off-policy, in which clean trajectories are collected without Herald in the loop, after which synthetic impairments are applied offline to train Herald and its verifier against packet loss and message manipulation, respectively. 
Stage 2 is on-policy, in which trajectories are collected with Herald active under online packet loss and adversarial attack, so that its own reconstructions and screening decisions feed back into the trajectories.
These trajectories are then mixed with the off-policy ones and distilled under the same objective, thereby reducing deployment-time covariate shift while mitigating catastrophic forgetting.

Importantly, Herald only verifies and reconstructs messages, without changing the agents' original policies or the parameters of the underlying MAS.
It introduces no additional inter-agent communications and thus no communication overhead.
When the channel is free of packet loss and adversarial attacks, Herald forwards every message unchanged, so the MAS behaves the same as originally trained.\looseness=-1\relax

The main contributions are summarized as follows:
\begin{itemize}[leftmargin=*]
\item We investigate the practical problem of retrofitting a communicative MAS trained under ideal communication with resilience to extreme packet loss and adversarial attacks, and propose Herald, a lightweight receiver-side module to enable it without any costly and unstable MARL retraining.\looseness=-1\relax
\item We build Herald around a Mamba-3-based latent-dynamics tracker, which keeps reconstruction accurate over long outages, equip it with a verifier that screens adversarially manipulated messages, and propose a two-stage distillation algorithm to train them efficiently without joint optimization with the MAS.
\item Extensive simulations on the StarCraft Multi-Agent Challenge (SMAC)~\cite{samvelyan2019smac} demonstrate that Herald recovers $92.1\%$ of the ideal win rate under $90\%$ packet loss, significantly outperforms multiple benchmarks and baselines, generalizes to unseen packet loss rates and bursty channels, and remains robust under multiple types of adversarial attacks.
\end{itemize}

The rest of the paper is organized as follows.
Sec.~\ref{sec:preliminaries} presents the preliminaries, Sec.~\ref{sec:Herald} details Herald, Sec.~\ref{sec:setup} describes the experiment setup, Sec.~\ref{sec:evaluation} reports the evaluation results, and Sec.~\ref{sec:conclusion} concludes the paper.

%% file: sections/3_preliminaries.tex
\section{Preliminaries}
\label{sec:preliminaries}
In this section, we first formulate the cooperative MAS as a partially observable Markov game (POMG).
Based on that, we review world-model-based communicative MARL~\cite{Egorov2022MAMBA} as the paradigm through which a cooperative MAS is obtained, and detail the execution pipeline of the resulting system.
Finally, we model the harsh communication environment in which such a system is deployed, i.e., the packet-loss channel over which its messages are exchanged.
Herald will be designed to retrofit such a system with resilience to extreme packet loss.\looseness=-1\relax

\subsection{Cooperative MAS as POMG}
\label{subsec:markov_game}

Consider an MAS consisting of a set of $N$ cooperative agents, indexed by $\mathcal{N}=\{1,\cdots,N\}$.
Their interaction with the environment is modeled as a POMG, defined as 
\begin{align*}
\mathcal{G}
=
\Big(
\mathcal{N},
\mathcal{S},
\{\mathcal{A}_i\}_{i\in\mathcal{N}},
\{\mathcal{O}_i\}_{i\in\mathcal{N}},
W,
\{\Omega_i\}_{i\in\mathcal{N}},
R,
\rho_0,
\gamma
\Big).
\end{align*}

Before explaining the elements in $\mathcal G$, we first define some short-hand notations: $\Delta(\mathcal{X})$ denotes the set of probability distributions over a set $\mathcal{X}$, and boldface symbols represent joint variables across all agents; for example, $\boldsymbol{\mathcal{A}}=\mathcal{A}_1\times\cdots\times\mathcal{A}_N$ denotes the joint action space, and $\boldsymbol{a}_t=(a_t^1,\ldots,a_t^N)\in\boldsymbol{\mathcal{A}}$ denotes the joint action taken at time step $t$.
Then, $\mathcal{S}$ denotes the global state space, while $\mathcal{A}_i$ and $\mathcal{O}_i$ denote the action and observation spaces of agent $i$, respectively.
The state-transition kernel $W:\mathcal{S}\times\boldsymbol{\mathcal{A}}\rightarrow\Delta(\mathcal{S})$ governs the evolution of the environment, i.e., $s_{t+1}\sim W\!\left(\cdot\mid s_t,\boldsymbol{a}_t\right)$.
The observation kernel $\Omega_i:\mathcal{S}\rightarrow\Delta(\mathcal{O}_i)$ determines the local observation of agent $i$, i.e., $o_t^i\sim\Omega_i(\cdot\mid s_t)$.
Moreover, $R:\mathcal{S}\times\boldsymbol{\mathcal{A}}\rightarrow\mathbb{R}$ is the team reward function shared by all agents, $\rho_0\in\Delta(\mathcal{S})$ is the initial-state distribution, and $\gamma\in[0,1)$ is the discount factor.

Due to partial observability, no agent can condition its action on the global state $s_t$, and must instead act on the information locally available to it, namely its own observation history and whatever it receives from its teammates.
Denoting this local information of agent $i$ at time step $t$ by $\tau_t^i$, the policy of agent $i$ reads $a_t^i\sim\pi_i(\cdot\mid\tau_t^i)$.
In the fully cooperative setting considered in this work, all agents seek to maximize the same expected discounted return
\begin{align*}
J(\boldsymbol{\pi})
=
\mathbb{E}_{\rho_0,W,\boldsymbol{\pi}}
\left[
\sum_{t=0}^{\infty}
\gamma^t
R(s_t,\boldsymbol{a}_t)
\right],
\end{align*}
where $\boldsymbol{\pi}=(\pi_1,\ldots,\pi_N)$ denotes the joint policy; $R$ and $\gamma$ are defined above in the POMG tuple $\mathcal{G}$.
We note that the concrete form of $R(s_t,\boldsymbol{a}_t)$ depends on the task of the MAS.

\subsection{World-Model-Based Communicative MARL}
\label{subsec:wm_comm_marl}
Obtaining a joint policy that maximizes $J(\boldsymbol{\pi})$ in the POMG of Sec.~\ref{subsec:markov_game} is the task of MARL, which can be broadly divided into model-free and model-based approaches.
The model-based ones improve sample efficiency by letting each agent learn a world model that predicts how its state evolves, and then training the policy on rollouts imagined by this model instead of on costly environment interactions.
However, a world model driven by the local information $o_t^i$ alone cannot anticipate the behaviors of the teammates, on which cooperation depends.
This motivates world-model-based communicative MARL, where the agents exchange messages so that every world model is also informed by its teammates.

A representative framework of this kind~\cite{Egorov2022MAMBA} integrates a communication module into the world model, enabling each agent to maintain a hidden world state from its local observations and the exchanged messages.
This hidden world state is a compact and fixed-size summary of the locally accessible information $\tau_t^i$, on which the policy of agent $i$ acts in place of $o_t^i$ itself.
Built on the recurrent state-space model (RSSM) of DreamerV2~\cite{hafner2020mastering}, it consists of a world model, representation model, and actor with parameters $\Phi\triangleq\{\phi,\varphi, \psi\}$. These parameters are shared across agents.
The execution pipeline of this framework can be briefly summarized below.\looseness=-1\relax

At time step $t-1$, each agent $i$ maintains a deterministic recurrent hidden state $h_{t-1}^i$.
At the beginning of $t$, every agent broadcasts its latent state $z_{t-1}^i$ together with its executed action $a_{t-1}^i$. 
Under ideal communication, the collections of latent states and actions, denoted by $\boldsymbol{z}_{t-1}$ and $\boldsymbol{a}_{t-1}$, are therefore available to every agent.
At every agent from $1$ to $N$, a Transformer-based communication module aggregates the received messages and produces a team-state embedding for each agent at time step $t$, i.e.,
\begin{align}
\big(e_t^1,\ldots,e_t^N\big)
=
\mathrm{Transformer}
\big(\boldsymbol{z}_{t-1},\boldsymbol{a}_{t-1}\big).
\label{eq:comm}
\end{align}

More specifically, in this module, the message $(z_{t-1}^j,a_{t-1}^j)$ of each agent $j$ is embedded as a single token in the Transformer.
When agent $i$ forms its team-state embedding $e_t^i$, it weights the token of every sender $j$ by a softmax over per-sender attention scores.

The $i$-th embedding $e_t^i$ is then used to update the recurrent hidden state of agent $i$ as
\begin{align}
h_t^i
=
f_{\phi}\!\left(h_{t-1}^i,e_t^i\right).
\label{eq:recurrent}
\end{align}

Given the updated hidden state $h_t^i$ and the current local observation $o_t^i$, the representation model can infer a latent state
\begin{align}
z_t^i
\sim
q_{\varphi}\!\left(
z_t^i
\,\middle|\,
h_t^i,o_t^i
\right).
\label{eq:posterior}
\end{align}
Here, $z_t^i$ is a discrete latent variable as in~\cite{hafner2020mastering,Egorov2022MAMBA}, represented by $\kappa$ categorical factors that are sampled independently given $(h_t^i,o_t^i)$.
Finally, the actor selects the action by
\begin{align}
a_t^i
\sim
\pi_{\psi}
\left(
a_t^i
\mid
h_t^i,z_t^i
\right),
\label{eq:actor}
\end{align}
after which the newly generated pair $(z_t^i,a_t^i)$ is broadcast to the other agents and used for communication at the next time step $t+1$, as formalized in Sec.~\ref{subsec:packet_loss}.

Executing \eqref{eq:comm}--\eqref{eq:actor} requires training the communication module, the recurrent model $f_{\phi}$, the representation model $q_{\varphi}$, and the actor $\pi_{\psi}$, which are optimized jointly by the world-model-based MARL.
Notably, world-model-based MARL is characterized by its use of imagined rollouts for policy training, which are generated by the world model itself rather than by interacting with the environment.
No real observation is therefore available along such a rollout, and the latent state can no longer be inferred from \eqref{eq:posterior}.
The framework therefore additionally learns a transition model $\hat z_t^i\sim p_{\phi}\big(\cdot\mid h_t^i\big)$, which predicts the latent state from the hidden state alone.
We refer readers to~\cite{Egorov2022MAMBA} for the complete training procedure.

Overall, this framework follows the centralized-training decentralized-execution (CTDE) paradigm: the parameters $\Phi$ are optimized jointly during training, whereas at execution each agent runs \eqref{eq:comm}--\eqref{eq:actor} on board using only its own observation and the messages received from its teammates, without access to the global state $s_t$.
The broadcast messages are therefore the only path through which teammates influence one another at execution time, which is exactly what packet loss disrupts and what Herald is designed to restore.

\subsection{Packet-Loss Channel in Harsh Environments}
\label{subsec:packet_loss}
The execution pipeline in Sec.~\ref{subsec:wm_comm_marl} is derived under the implicit assumption that every broadcast message reaches every teammate. 
This assumption is untenable in harsh environments, such as underwater, underground, and post-disaster deployments, where severe attenuation, multipath, blockage, and interference make each link unreliable, and agent mobility further renders the connectivity intermittent.

To facilitate presentation, we adopt a general channel model that is agnostic to the physical cause of the impairment. 
Such a model is coherent with how messages are delivered in practical digital communication systems, where messages are carried in packets and every packet has an error-detecting code, e.g., a cyclic redundancy check~\cite{Graf2024ChannelFEC,Graf2024ErrorPatterns}.
A packet whose check fails is discarded rather than passed to the upper layer.
Consequently, whatever the underlying cause of the impairment, a message is affected at the granularity of a packet: the receiver either obtains the message intact or obtains no message.

Specifically, we consider a time-slotted communication system, in which time is divided into slots of equal duration synchronized across agents by their local clocks.
Following the causal update schedule of Sec.~\ref{subsec:wm_comm_marl}, at the beginning of time step $t$ agent $j$ packetizes the latent state and the action it produced at the preceding step into the message
\begin{align}
m_t^j
\triangleq
\left(z_{t-1}^j,a_{t-1}^j\right),
\label{eq:Herald_message}
\end{align}
and broadcasts it to the other agents.
Since the latent state $z_{t-1}^j$ and the action $a_{t-1}^j$ are categorical variables, which are kept low-dimensional to facilitate the MARL~\cite{hafner2020mastering,Egorov2022MAMBA}, we assume that each message $m_t^j$ is carried by a single packet, so that a packet loss removes the latent state and the action of that agent jointly rather than degrading them partially.

For each directed teammate link $j\rightarrow i$, we define the \emph{delivery indicator} as
\begin{align}\label{equ_delivery_ind}
\delta_t^{j\rightarrow i}
=
\begin{cases}
1, & \text{if $m_t^j$ is received},\\
0, & \text{if $m_t^j$ is lost}.
\end{cases}
\end{align}
Since the self-message is always available to each agent $i$, we have $\delta_t^{i\rightarrow i}=1,\ \forall i,t$.

As a typical instance of {packet-loss channel}, we take packet delivery on different directed teammate links to be independently distributed as
\begin{align}
\delta_t^{j\rightarrow i}
\sim
\operatorname{Bernoulli}(1-p),
\qquad j\neq i,
\label{eq:Herald_bernoulli_channel}
\end{align}
where $p\in[0,1]$ is the packet-loss rate.\!\footnote{Temporally correlated bursty losses, typical of mobility-induced blockage and shadowing, are also evaluated in Sec.~\ref{sec:evaluation}.}
We note that $p$ corresponds to the \emph{residual} packet loss rate after whatever link-layer error-correction and retransmission the underlying protocol provides. 

It is worthwhile mentioning that the messages in \eqref{eq:Herald_message} are consumed by a real-time decision loop, which bounds the number of attempts that fit within one control step and renders the loss unrecoverable once that budget is exhausted.
Therefore, $p$ can remain substantial even in the presence of link-layer retransmission. 
Moreover, in this paper, we focus on the harsh environment with potentially extreme packet loss due to severe interference, considering $p\ge0.6$ reaching above $0.9$. 
Under such conditions, an agent is usually deprived of a teammate's message over a long sequence of steps: Based on \eqref{eq:Herald_bernoulli_channel}, a nonempty burst of consecutive losses has mean length $1/(1-p)$ steps for $p<1$, and communication vanishes as $p\rightarrow1$.
Buffering the last received message or waiting for a delayed one is of little use across such prolonged outages, which is precisely the gap that Herald is designed to fill.

%% file: sections/4_HERALD.tex
\section{Design and Training of Herald}
\label{sec:Herald}
\begin{figure*}[t]
\centering
\includegraphics[width=0.95\textwidth]{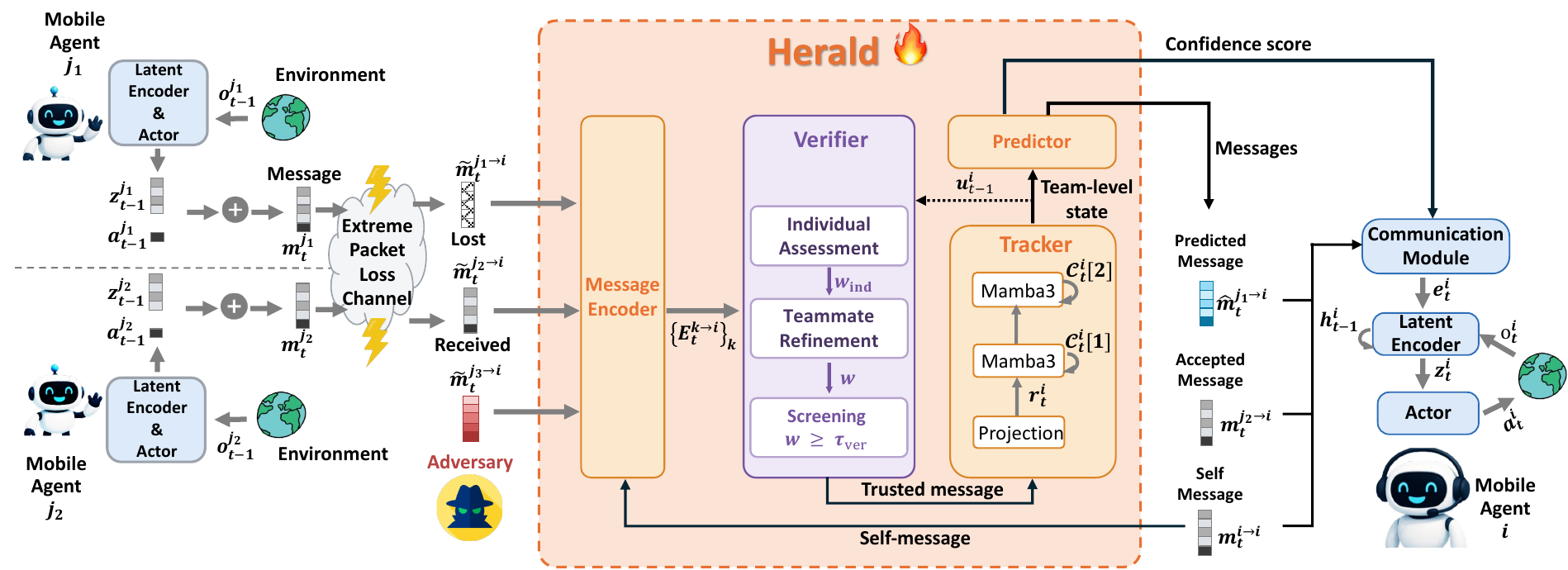}
\caption{Processing pipeline of Herald at one step.}
\label{fig:herald_execution}
\vspace{-1em}
\end{figure*}
In this section, we present Herald, a lightweight receiver-side module that retrofits a trained communicative MAS with resilience to extreme packet loss and adversarial messages.
Metaphorically, this module works like an intelligent herald employed by each agent to facilitate the collaboration with other teammates over a lossy channel.
It relays every accepted message faithfully and, whenever a message is lost or seems adversarial, completes it by prediction from its record of past communication and from the messages it has accepted.
It further reports candidly how much confidence it places in each prediction, so that its agent can weigh the completed message accordingly, thereby facilitating collaboration within the agent team under unreliable communication.\looseness=-1\relax

Below, we first elaborate the module design of Herald in Sec.~\ref{subsec:Herald_architecture}, and then present its training algorithm in Sec.~\ref{subsec:Herald_training}, which distills Herald from the frozen backbone MAS in two stages. Throughout, all quantities are indexed by the step in which the corresponding message is transmitted and consumed, following Sec.~\ref{subsec:packet_loss}.

\subsection{Module Design of Herald}
\label{subsec:Herald_architecture}
Herald operates independently at each receiver and consists of four main components: a message encoder, a verifier, a tracker, and a predictor.
We first describe the message encoder, tracker, and predictor, which together form the message reconstruction pipeline.
We then introduce the verifier for adversarial environments.
It assesses the trustworthiness of received messages and determines whether to accept them. Rejected messages are treated as missing and recovered through the same reconstruction pipeline.
The interconnection among the components is illustrated in Fig.~\ref{fig:herald_execution}.\looseness=-1\relax

\subsubsection{Message Encoder}
To facilitate description, we first consider the non-adversarial setting, where messages are subject only to packet loss and every received message is authentic.
For every teammate transmitter~(Tx) agent $j$, the message encoder encodes the message observed on link $j\rightarrow i$, together with its delivery status, the identity of that teammate, and how long it has been lost, into a single message embedding.
For each incoming link, the gap is expressed as
\begin{align}
g_t^{j\rightarrow i}
=
\begin{cases}
0, & \delta_t^{j\rightarrow i}=1,\\
g_{t-1}^{j\rightarrow i}+1, & \delta_t^{j\rightarrow i}=0,
\end{cases}
\label{eq:Herald_gap}
\end{align}
which counts the consecutive steps elapsed since the last successful reception from sender $j$, with $g_0^{j\rightarrow i}=0$ and $g_t^{i\rightarrow i}=0$.
For each communication link $j\rightarrow i$, Herald constructs a message embedding by
\begin{align}
E_t^{j\rightarrow i}
=
\mathrm{MsgEnc}_{\theta}
\left(
\widetilde m_t^{j\rightarrow i},
\delta_t^{j\rightarrow i},
g_t^{j\rightarrow i},
d_j
\right),
\label{eq:Herald_slot_encoding}
\end{align}
where $\mathrm{MsgEnc}_{\theta}$ is the encoding function parameterized by parameters $\theta$, $d_j$ denotes a learnable sender-identity embedding, and $\widetilde m_t^{j\rightarrow i}$ denotes the \emph{observed} message content on link $j\rightarrow i$, defined as
\begin{align}
\widetilde m_t^{j\rightarrow i}
\triangleq
\begin{cases}
m_t^{j},
& \delta_t^{j\rightarrow i}=1,\\
m_{\varnothing}=\left(z_{\varnothing},a_{\varnothing}\right),
& \delta_t^{j\rightarrow i}=0.
\end{cases}
\label{eq:Herald_observed_message}
\end{align}
Here $m_t^{j}$ is the transmitted message in \eqref{eq:Herald_message}, whose latent-state and action fields are passed to the message encoder verbatim whenever the packet is delivered. Otherwise, both fields are replaced by the placeholder $m_{\varnothing}$, whose components $z_{\varnothing}$ and $a_{\varnothing}$ are learnable missing-message embeddings optimized jointly with the remaining Herald parameters. 

\subsubsection{Tracker}\label{s3ec_tracker}
The tracker is responsible for capturing the latent dynamics, maintaining them as a temporally evolving summary of the team-level state observed by each receiver.
For Rx agent $i$, the complete set of message embeddings is first concatenated in a fixed sender order and projected into a receiver-level representation: $r_t^i=W_r\left[E_t^{1\rightarrow i};\ldots;E_t^{N\rightarrow i}\right]+b_{\mathrm{proj}}$, where $W_r$ and $b_{\mathrm{proj}}$ are a learned projection matrix and bias.
Representation $r_t^i$ is then handled by the tracker.
Although the parameters of the trackers are shared across agents, each agent maintains an independent internal recurrent state, which is calculated by
\begin{align}
\left(
u_t^i,
\mathcal{C}_t^i
\right)
=
\mathrm{Tracker}_{\theta}
\left(
r_t^i;
\mathcal{C}_{t-1}^i
\right),
\label{eq:Herald_tracker}
\end{align}
where $u_t^i$ denotes the team-level state produced at step $t$, 
and $\mathcal{C}_t^i$ denotes the internal recurrent state of the tracker.

We design the tracker with Mamba-3 blocks~\cite{lahoti2026mamba3} for three reasons.
First, it compresses the past into a fixed-size state, so its per-step cost stays constant.
Second, its input-dependent retention factor preserves evidence over many steps, which the reconstruction relies on at high packet-loss rates.
Finally, its complex-valued state transitions and exponential-trapezoidal recurrence model the latent dynamics more faithfully than conventional recurrent neural networks.\looseness=-1\relax

We next describe one Mamba-3 block, omitting the receiver and block indices for clarity.
Each block carries a memory matrix $M_t$ and maps its input $x_t^{\mathrm{in}}$ to
\begin{align}
x_t^{\mathrm{out}}
=
x_t^{\mathrm{in}}
+
W_{\mathrm{out}}
\left[
\left(
M_tQ_t
+
DV_t
\right)
\odot
\operatorname{SiLU}
\left(
\Gamma_t
\right)
\right],
\label{eq:Herald_mamba_block_output}
\end{align}
where $\odot$ is the element-wise product and $W_{\mathrm{out}}$ is the output projection.
Reading~\eqref{eq:Herald_mamba_block_output} from the inside out, the term $M_tQ_t$ retrieves the accumulated history along a read direction $Q_t$; the term $DV_t$ passes through the current evidence $V_t$ directly with a learned weight $D$; the gate $\operatorname{SiLU}(\Gamma_t)$, formed from a gating feature $\Gamma_t$, selects how much of the result is exposed to the next block; and the term $x_t^{\mathrm{in}}$ is from a residual connection preserving the input.
Apart from the memory $M_t$, every quantity in~\eqref{eq:Herald_mamba_block_output} is obtained from $x_t^{\mathrm{in}}$ through a single input projection $W_{\mathrm{in}}$.
This projection also produces the remaining input-dependent quantities introduced below.\looseness=-1\relax

For Herald, two core mechanisms of Mamba-3 distinguish it from a conventional gated recurrence, both of which concern the evolution of $M_t$.
The first, termed \emph{exponential-trapezoidal discretization}, governs how the continuously evolving evidence, which is observed at discrete steps, is accumulated into the memory.
Denoting $K_t$ for the write direction paired with $V_t$, it yields the recurrence:
\begin{align}
M_t
=
\lambda_t^{\mathrm{ret}}
M_{t-1}
+
\omega_{t,-}
V_{t-1}K_{t-1}^{\mathsf T}
+
\omega_{t,+}
V_tK_t^{\mathsf T}.
\label{eq:Herald_mamba_state_update}
\end{align}

In~\eqref{eq:Herald_mamba_state_update}, the first term carries the existing memory forward, discounted by the retention factor $\lambda_t^{\mathrm{ret}}=\exp\left(A_t\Delta_t^{\mathrm{ssm}}\right)\in(0,1)$, where $A_t<0$ is a learned decay rate and $\Delta_t^{\mathrm{ssm}}>0$ is a learned step size.
Here, $\Delta_t^{\mathrm{ssm}}$ represents a learned interval, over which the underlying dynamics are assumed to evolve continuously and the memory decays accordingly.
The remaining two terms in~\eqref{eq:Herald_mamba_state_update} write new evidence into the memory, each depositing a value along its paired write direction through an outer product.
They correspond to the two endpoints of the interval, with $V_{t-1}K_{t-1}^{\mathsf T}$ being the evidence observed at its start and $V_tK_t^{\mathsf T}$ that observed at its end.
Their weights $\omega_{t,-}=(1-\eta_t)\Delta_t^{\mathrm{ssm}}\lambda_t^{\mathrm{ret}}$ and $\omega_{t,+}=\eta_t\Delta_t^{\mathrm{ssm}}$ follow the trapezoidal rule, both scaling with the interval length $\Delta_t^{\mathrm{ssm}}$ and splitting the update according to a balance $\eta_t\in(0,1)$, with $A_t$, $\Delta_t^{\mathrm{ssm}}$, and $\eta_t$ all produced by projecting $x_t^{\mathrm{in}}$ with $W_{\mathrm{in}}$.
The weight $\omega_{t,-}$ carries the additional factor $\lambda_t^{\mathrm{ret}}$, as its contribution also decays while traversing the interval.
A gated recurrent unit~(GRU) can only overwrite or forget its state using the evidence at a single instant, whereas~\eqref{eq:Herald_mamba_state_update} integrates two consecutive observations within one transition.
It thereby approximates the underlying continuous dynamics to second order, giving a more faithful account of how a team-level state moves between two steps.\looseness=-1\relax

The second, namely the \emph{complex-valued state space}, governs how the stored evidence evolves and is retrieved.
As $\lambda_t^{\mathrm{ret}}$ is a positive scalar, decay alone can only shrink what is stored.
Making the state complex-valued instead amounts to rotating its coordinates by a cumulative phase $\Theta_t=\Theta_{t-1}+\uppi\tanh(\zeta_t)\Delta_t^{\mathrm{ssm}}$, where the increment $\zeta_t$ is another output of the input projection with $W_{\mathrm{in}}$ and $\uppi$ is the circle constant.
The projection with $W_{\mathrm{in}}$ also produces a write feature $B_t$ and a read feature $C_t$.
Applying the same phase to both gives the write and read directions $K_t=\operatorname{RoPE}(B_t,\Theta_t)$ and $Q_t=\operatorname{RoPE}(C_t,\Theta_t)$, where the rotary position embedding~(RoPE)~\cite{su2024roformer} treats each pair of adjacent coordinates as a complex number and multiplies the $n$-th of them by $e^{\mathrm{j}\Theta_{t,n}}$, thereby rotating that coordinate pair by the angle $\Theta_{t,n}$.

Rotating the state in this manner benefits the tracker in two respects.
First, stored state may oscillate instead of only fading.
Second, the alignment between read and write directions depends on their relative phase, so the memory records how long ago each piece of evidence arrived.

In summary, internal recurrent state $\mathcal{C}_t^i$ in~\eqref{eq:Herald_tracker} consists of $M_t^i\,,\Theta_t^i\,,V_t^i$, and $K_t^i$.
As for $u_t^i$, it is obtained by feeding the output of one block into the next and applying a normalization to the output of the last block.

\subsubsection{Predictor}\label{s3ec_predictor}
The predictor of Rx agent $i$ reconstructs each lost message by using $u_t^i$ as well as the message embeddings of the arrived teammate packets.
Specifically, for each Tx agent $j$ whose packet is lost, it first constructs a query $q_t^{j\rightarrow i}=E_t^{j\rightarrow i}+W_u u_t^i$. 
The query then attends to all message embeddings, i.e., $\{E_t^{k\rightarrow i}\}_{k\in\mathcal{N}}$, allowing the prediction to exploit both temporal context and the cross-agent information available.\looseness=-1\relax

Based on the resulting representation, the predictor outputs the reconstructed latent state, the reconstructed action, and a confidence score:
\begin{align}
\left(
\widehat z_{t-1}^{j\rightarrow i},
\widehat a_{t-1}^{j\rightarrow i},
c_t^{j\rightarrow i}
\right)
=
\mathrm{Pred}_{\theta}
\left(
q_t^{j\rightarrow i},
\left\{
E_t^{k\rightarrow i}
\right\}_{k\in\mathcal{N}}
\right),
\end{align}
where $c_t^{j\rightarrow i}\in(0,1]$ quantifies the predicted reliability of the reconstruction.
Both reconstructions are produced in categorical form by a softmax, applied independently to each of the $\kappa$ latent factors for $\widehat z_{t-1}^{j\rightarrow i}$ and over the action space for $\widehat a_{t-1}^{j\rightarrow i}$.
They are therefore soft distributions over the message alphabet rather than hard samples.
The reconstructed message is then defined as $\widehat m_t^{j\rightarrow i}=(\widehat z_{t-1}^{j\rightarrow i},\widehat a_{t-1}^{j\rightarrow i})$.

In the non-adversarial setting, Herald modifies only the messages affected by packet loss.
For each link $j\rightarrow i$, the message supplied to the communication module defined in~\eqref{eq:comm} is
\begin{align}
\overline m_t^{j\rightarrow i}
=
\begin{cases}
m_t^{j},
& \delta_t^{j\rightarrow i}=1,\\
\widehat m_t^{j\rightarrow i},
& \delta_t^{j\rightarrow i}=0.
\end{cases}
\label{eq:Herald_repaired_message}
\end{align}
Hence, received messages are forwarded unchanged, whereas lost messages are replaced by their predictor-generated soft reconstructions.

Reconstructed messages are, however, not equally reliable, and Herald exploits the predicted confidence to regulate how much each of them is relied upon.
The place to do so is the inter-agent attention of the frozen communication module, which, as described in Sec.~\ref{subsec:wm_comm_marl}, is where the messages from different Tx agents are integrated to form $e_t^i$.
Here, the predicted confidence is first converted into an additive bias by
\begin{align}
b_t^{j\rightarrow i}
=
\left(
1-\delta_t^{j\rightarrow i}
\right)
\log
\left(
\max\{c_t^{j\rightarrow i},\epsilon\}
\right),
\label{eq:Herald_confidence_bias}
\end{align}
where $0<\epsilon<1$ is a small constant introduced for numerical stability. A received message therefore incurs no bias, while a reconstructed message is assigned a non-positive bias whose magnitude increases as its confidence decreases.

Without loss of generality, we assume each message $\overline m_t^{j\rightarrow i}$ is embedded into a token $\chi_t^{j\rightarrow i}$ in the Transformer of the communication module, and the bias is added to its attention score before the softmax, i.e.,
\begin{align}
\alpha_t^{j\rightarrow i}
=
\operatorname{softmax}_{j}
\left(
\frac{\left(W_q\chi_t^{i\rightarrow i}\right)^{\mathsf T}W_k\chi_t^{j\rightarrow i}}{\sqrt{d}}
+
b_t^{j\rightarrow i}
\right),
\label{eq:Herald_confidence_attention}
\end{align}
where $W_q$ and $W_k$ are the frozen query and key projections, $d$ is the key dimension, and $\chi_t^{i\rightarrow i}$ and $\chi_t^{j\rightarrow i}$ denote the tokens of $\overline m_t^{i\rightarrow i}$ and $\overline m_t^{j\rightarrow i}$, respectively.
When $c_t^{j\rightarrow i}\rightarrow1$, the reconstruction is treated as if it were a received message, whereas when $c_t^{j\rightarrow i}\rightarrow0$, it is effectively masked out.
We apply the same bias to every attention layer of the communication module.
The resulting communication embedding $e_t^i$ is then passed to the original frozen execution pipeline of the MAS according to Eqs.~\eqref{eq:recurrent}--\eqref{eq:actor}.

\emph{\textbf{Remark}}:
Herald becomes transparent to the underlying MAS if the network environment recovers from severe interference, as a received message is forwarded exactly by~\eqref{eq:Herald_repaired_message} and receives a zero bias by~\eqref{eq:Herald_confidence_bias}.
Consequently, adopting Herald does not risk degrading the performance that the MAS attains under benign channels, and no switching logic is needed to disable it once packet loss ceases.

\subsubsection{Verifier}
\label{subsec:Herald_verifier}
We now extend Herald to adversarial environments, where a teammate's message may have been manipulated by an adversary, either in flight or at a hijacked sender.
If such a message is accepted, it may mislead the agent's current decision and distort the tracker's history, affecting future message reconstructions.
Therefore, we design a verifier for Herald to assess received messages before the tracker updates its history.
The verifier first assesses each message individually against a reference constructed from the preceding team-state summary and the receiver's local self-message, and then refines this assessment using other messages received at the same step.

\textbf{History-Based Individual Assessment}: To determine whether a received message has been adversarially manipulated, the verifier first evaluates its consistency with the receiver's communication history and local information.
Specifically, to assess a message from sender $j$, receiver $i$ combines its preceding team-state summary $u_{t-1}^i$, local self-message embedding $E_t^{i\rightarrow i}$, sender identity $d_j$, and elapsed interval $g_{\mathrm{acc},t-1}^{j\rightarrow i}$ into a history-context vector
\begin{align}
\boldsymbol h_{\mathrm{ver},t}^{j\rightarrow i}
&=\mathrm{Context}\!\left(
u_{t-1}^i,E_t^{i\rightarrow i},d_j,
g_{\mathrm{acc},t-1}^{j\rightarrow i}\right),
\label{eq:Herald_verifier_context}
\end{align}
where $g_{\mathrm{acc},t-1}^{j\rightarrow i}$ counts the consecutive steps without an accepted message from $j$ up to step $t-1$.
The local self-message is assumed authentic.
Then, a reference constructor maps $\boldsymbol h_{\mathrm{ver},t}^{j\rightarrow i}$ to the predictive distributions $\boldsymbol p_{z,t}^{\mathrm{ref},j\rightarrow i}$ and $\boldsymbol p_{a,t}^{\mathrm{ref},j\rightarrow i}$ for the message's latent state and action, respectively, as
\begin{align}
\left(\boldsymbol p_{z,t}^{\mathrm{ref},j\rightarrow i},
\boldsymbol p_{a,t}^{\mathrm{ref},j\rightarrow i}\right)
&=\mathrm{Ref}\!\left(
\boldsymbol h_{\mathrm{ver},t}^{j\rightarrow i}\right),
\label{eq:Herald_verifier_prior}
\end{align}
where $\boldsymbol p_{z,t}^{\mathrm{ref},j\rightarrow i}$ comprises one categorical distribution for each of the $\kappa$ latent factors, and $\boldsymbol p_{a,t}^{\mathrm{ref},j\rightarrow i}$ is a categorical distribution over $\mathcal A_j$.
These distributions describe the expected contents of sender $j$'s message at step $t$ and provide a reference for assessing the received message.

A discrepancy between the received message and its reference may indicate adversarial manipulation, but it may also arise because the available history does not support a prediction.
The verifier therefore measures both the discrepancy between the received and predicted contents and the uncertainty of the prediction itself.
For the received message from sender $j$, let $\widetilde{\boldsymbol z}_{t-1,f}^{j\rightarrow i}$ and $\widetilde{\boldsymbol a}_{t-1}^{j\rightarrow i}$ denote its $f$-th latent-state factor and action field, respectively.
The tilde distinguishes the received fields, which may have been manipulated, from the sender's original contents.
Their cross-entropies under the reference distributions give the discrepancy measures $S_{z,t}^{j\rightarrow i}$ and $S_{a,t}^{j\rightarrow i}$, while the entropies $H_{z,t}^{j\rightarrow i}$ and $H_{a,t}^{j\rightarrow i}$ quantify the uncertainty of these reference distributions, which are calculated as
\begin{subequations}\label{eq:Herald_verifier_statistics}
\begin{align}
S_{z,t}^{j\rightarrow i}
&=\frac{1}{\kappa}\sum_{f=1}^{\kappa}\mathrm{CE}\!\left(
\widetilde{\boldsymbol z}_{t-1,f}^{j\rightarrow i},
\boldsymbol p_{z,t,f}^{\mathrm{ref},j\rightarrow i}\right),
\\
S_{a,t}^{j\rightarrow i}
&=\mathrm{CE}\!\left(\widetilde{\boldsymbol a}_{t-1}^{j\rightarrow i},
\boldsymbol p_{a,t}^{\mathrm{ref},j\rightarrow i}\right),
\\
H_{z,t}^{j\rightarrow i}
&=\frac{1}{\kappa}\sum_{f=1}^{\kappa}\mathrm{H}\!\left(
\boldsymbol p_{z,t,f}^{\mathrm{ref},j\rightarrow i}\right),
\\
H_{a,t}^{j\rightarrow i}
&=\mathrm{H}\!\left(\boldsymbol p_{a,t}^{\mathrm{ref},j\rightarrow i}\right),
\end{align}
\end{subequations}
where $\mathrm{CE}$ and $\mathrm{H}$ denote cross-entropy and entropy, respectively, and $\boldsymbol p_{z,t,f}^{\mathrm{ref},j\rightarrow i}$ is the categorical distribution for latent factor $f\in\{1,\ldots,\kappa\}$.

Beyond using the above metrics as evidence for the assessment, the verifier also leverages the context vector and the received message embedding, which are combined into a feature vector, i.e., 
$\boldsymbol x_{\mathrm{ver},t}^{j\rightarrow i}
=[\boldsymbol h_{\mathrm{ver},t}^{j\rightarrow i},
E_t^{j\rightarrow i} ]$, where embedding $E_t^{j\rightarrow i}$ is computed from the received message using the message encoder in~\eqref{eq:Herald_slot_encoding}. 
The verifier then feeds this feature vector together with the four metrics into a classifier $\mathrm{Cls}(\cdot)$, which first produces an intermediate logit $\ell_{\mathrm{ind},t}^{j\rightarrow i}\in\mathbb R$ and then derives a trustworthiness score $w_{\mathrm{ind},t}^{j\rightarrow i}\in(0,1)$ by using a sigmoid layer, which is expressed by
\begin{align}
w_{\mathrm{ind},t}^{j\rightarrow i}
& =\mathrm{Cls}(\boldsymbol x_{\mathrm{ver},t}^{j\rightarrow i},S_{z,t}^{j\rightarrow i}\!,S_{a,t}^{j\rightarrow i}\!,H_{z,t}^{j\rightarrow i}\!,H_{a,t}^{j\rightarrow i}) \label{eq:Herald_verifier_initial_score_1}
\\
& =\mathrm{Sigmoid}\!\left(\ell_{\mathrm{ind},t}^{j\rightarrow i}\right).
\label{eq:Herald_verifier_initial_score_2}
\end{align}

For the local message of agent $i$, we directly set $w_{\mathrm{ind},t}^{i\rightarrow i}=1$ as it is assumed authentic and free of manipulation.

\textbf{Teammate-Assisted Refinement}: Messages received from other teammates at the same step can help assess whether a candidate message is consistent with the current environment.
However, these messages may themselves have been manipulated.
Therefore, the verifier needs to aggregate evidence from teammates according to both its relevance to the candidate message and their individual trustworthiness.

To realize this aggregation, we use the attention mechanism~\cite{vaswani2017transformer} with a trustworthiness bias added to the relevance scores.
Following~\cite{vaswani2017transformer}, we project each received message's feature vector $\boldsymbol x_{\mathrm{ver},t}^{j\rightarrow i}$ into a query, a key, and value vectors.
Moreover, inspired by the key normalization in~\cite{govindarajan2026quest}, we normalize the Euclidean norms of queries and keys to prevent individual vectors from attracting excessive attention.
We also limit each value vector's norm at $V_{\mathrm{ver}}^{\max}>0$ to bound individual contribution.
For each sender $k$'s message ($k\neq j$), denoting the resulting vectors by ${\boldsymbol q_{t}^{k\rightarrow i}}$, ${\boldsymbol k_{t}^{k\rightarrow i}}$, and ${\boldsymbol v_{t}^{k\rightarrow i}}$, an attention head aggregates the teammate evidence for the sender $j$ by
\begin{align}
a_{t,k}^{j\rightarrow i}
&=\big({\boldsymbol q}_{t}^{j\rightarrow i}\big)^{\mathsf T}
{\boldsymbol k}_{t}^{k\rightarrow i}+\log\max\{w_{\mathrm{ind},t}^{k\rightarrow i},\epsilon\},
\label{eq:Herald_verifier_peer_evidence_1}
\\
\boldsymbol o_{t}^{j\rightarrow i}
&=\sum_{k\in \mathcal P_t^{j\rightarrow i}}
\frac{\exp(a_{t,k}^{j\rightarrow i})}
{1+\sum_{l\in \mathcal P_t^{j\rightarrow i}}\exp(a_{t,l}^{j\rightarrow i})}\, {\boldsymbol v}_{t}^{k\rightarrow i},
\label{eq:Herald_verifier_peer_evidence_2}
\end{align}
where 
$a_{t,k}^{j\rightarrow i}$ is the trustworthiness-aware attention score,
$\mathcal R_t^i=\{k\in\mathcal N\mid\delta_t^{k\rightarrow i}=1\}$ denotes the senders of all received messages, including the local self-message, 
and $\mathcal P_t^{j\rightarrow i}=\mathcal R_t^i\setminus\{j\}$ excludes the candidate to focus on teammate contribution.
The constant $1$ in the denominator allows the total weight assigned to teammate evidence to be small when all contributing messages receive low attention scores.

The aggregated evidence is used to refine the individual assessment.
Specifically, a learned function $\mathrm{TeamCorr}(\cdot)$ maps $\boldsymbol o_{t}^{j\rightarrow i}$ to an additive refinement value to the individual logit $\ell_{\mathrm{ind},t}^{j\rightarrow i}$.
Applying a sigmoid function to the corrected logit gives the final trustworthiness score
\begin{align}
w_t^{j\rightarrow i}
&=\mathrm{Sigmoid}\!\left(\ell_{\mathrm{ind},t}^{j\rightarrow i} + \mathrm{TeamCorr}(\boldsymbol o_{t}^{j\rightarrow i})\right).
\label{eq:Herald_verifier_final_score}
\end{align}

Herald accepts a received teammate message when $w_t^{j\rightarrow i}\ge\tau_{\mathrm{ver}}$, where $\tau_{\mathrm{ver}}\in(0,1)$ is a preset threshold, and otherwise discards its payload and treats the message as lost.
After this screening, accepted messages are passed to the tracker in Sec.~\ref{s3ec_tracker}, while the rejected messages are treated as lost.
The predictor in Sec.~\ref{s3ec_predictor} thus reconstructs both the lost and the rejected messages using the updated history and accepted messages.

\subsection{Training Algorithm for Herald}
\label{subsec:Herald_training}

In the following, we describe the training of Herald's lost-message prediction and adversarial-message verification capabilities in turn, presenting the objective function and corresponding procedures for each.

\subsubsection{Training the Prediction Capability}
\label{subsec:Herald_prediction_training}

To train Herald's prediction capability, we first formulate the objective function and then detail its training procedures, for which we propose a two-stage distillation method.

\textbf{Objective Function Formulation.}
The prediction capability of Herald is trained by distillation in a simulated environment, in which packet loss is emulated so that every lossy trajectory has a loss-free counterpart.
The teacher is the frozen MAS executed on the complete message set, whereas the student is Herald.
Specifically, Herald is distilled to match the teacher-side soft outputs rather than the discrete message, as well as the action distributions produced by their policy networks, and the resulting embeddings of the communication module.
More specifically, the Herald training objective consists of three terms:
\begin{align}
\mathcal{L}_{\mathrm{Herald}}
=
\mathcal{L}_{\mathrm{rec}}
+
\mathcal{L}_{\mathrm{dev}}
+
\beta\mathcal{L}_{\mathrm{cfd}},
\label{eq:Herald_total_loss}
\end{align}
where $\mathcal{L}_{\mathrm{rec}}$ penalizes the discrepancy between a reconstructed message and its teacher-side target, $\mathcal{L}_{\mathrm{dev}}$ penalizes the deviation the reconstruction induces downstream, and $\mathcal{L}_{\mathrm{cfd}}$ encourages the predicted confidence to reflect the actual reconstruction quality with $\beta$ being a weight hyperparameter. 
We elaborate on the three loss terms in turn.\looseness=-1\relax

\emph{Message Reconstruction Loss $\mathcal{L}_{\mathrm{rec}}$}:
For each missing link $j\rightarrow i$, Herald matches the predicted latent-state and action distributions to the corresponding soft targets produced by the frozen backbone MAS. Let $\boldsymbol p_{z,t}^{j,\star}$ and $\widehat{\boldsymbol p}_{z,t}^{j\rightarrow i}$ denote the target and predicted distributions of the latent factors in $m_t^j$, i.e., $z_{t-1}^j$, respectively, and let $\boldsymbol p_{a,t}^{j,\star}$ and $\widehat{\boldsymbol p}_{a,t}^{j\rightarrow i}$ denote the corresponding distributions of $a_{t-1}^j$. The message reconstruction loss is
\begin{align}
\mathcal{L}_{\mathrm{rec}}
&=
\frac{1}{|\mathcal{I}_{\mathrm{miss}}|}\sum_{(t,i,j)\in\mathcal{I}_{\mathrm{miss}}}\hspace{-2ex}
D_{\mathrm{KL}}
\left(
\boldsymbol p_{z,t}^{j,\star}, \boldsymbol p_{a,t}^{j,\star}
\middle\|
\widehat{\boldsymbol p}_{z,t}^{j\rightarrow i}, \widehat{\boldsymbol p}_{a,t}^{j\rightarrow i}
\right),
\label{equ:message_recon_loss}
\end{align}
where $\mathcal{I}_{\mathrm{miss}}=\{(t,i,j):j\neq i,\;\delta_t^{j\rightarrow i}=0\}$ indexes the emulated lost teammate messages by step, receiver, and sender, and $D_{\mathrm{KL}}\left(\cdot\middle\|\cdot\right)$ denotes the Kullback--Leibler divergence.

\emph{Team-State Deviation Loss $\mathcal{L}_{\mathrm{dev}}$}:
A small message-level reconstruction error does not necessarily translate into a small deviation downstream.
We therefore additionally align the team-state embeddings $e_{t}^i$ produced by the reconstructed and the original message, which are denoted by $\widehat{e}_{t}^i$ and $e_{t}^{\star,i}$, respectively. 
We define the team-state deviation loss as
\begin{align}
\mathcal{L}_{\mathrm{dev}}
=
\frac{1}{|\mathcal{I}_{\mathrm{dev}}|}
\sum_{(t,i)\in\mathcal{I}_{\mathrm{dev}}}
\left\|
\widehat{e}_{t}^i
-
e_{t}^{\star,i}
\right\|_2^2,
\end{align}
where $\mathcal{I}_{\mathrm{dev}}$ contains receiver steps with at least one missing teammate message. %

\emph{Confidence Prediction Loss $\mathcal{L}_{\mathrm{cfd}}$}:
Finally, the confidence prediction is trained to reflect the message reconstruction quality. 
For each missing message, we define the soft confidence target $c_t^{\star,\,j\rightarrow i}=\exp(-\ell_{\mathrm{rec},t}^{j\rightarrow i})$ where $\ell_{\mathrm{rec},t}^{j\rightarrow i}$ is the per-item message reconstruction loss summed in~\eqref{equ:message_recon_loss} for $(t,i,j)$.
Then, the confidence prediction loss is calculated by
\begin{align}
\mathcal{L}_{\mathrm{cfd}}
=
\frac{1}{|\mathcal{I}_{\mathrm{miss}}|}
\sum_{(t,i,j)\in\mathcal{I}_{\mathrm{miss}}}
\|
c_t^{j\rightarrow i}
-
\operatorname{sg}[c_t^{\star,\,j\rightarrow i}]
\|^2.
\end{align}
A more accurate reconstruction therefore receives a higher confidence target, while the stop-gradient operation, i.e., $\operatorname{sg}[\cdot]$, prevents gradients from flowing through the reconstruction-derived confidence target.

\textbf{Training Method.}
\label{subsec:Herald_two_stage_training}
We propose a \emph{two-stage distillation-based algorithm} to train the prediction capability of Herald efficiently.
It comprises two complementary stages, including an \emph{off-policy stage}, which supervises Herald on packet-loss-free trajectories collected without it in the loop, and an \emph{on-policy stage}, which trains it on the trajectories that its own message predictions induce.

\emph{Stage 1: Off-Policy Distillation.}
We first execute the MAS under lossless communication to collect an off-policy trajectory dataset $\mathcal{D}_{\mathrm{off}}$, together with the soft latent-state and action distributions used as reconstruction targets. During training, per-link packet losses are synthetically applied to these trajectories, and Herald is optimized using \eqref{eq:Herald_total_loss}.
This stage is necessary because an untrained Herald produces arbitrary reconstructions, which would immediately collapse the collaboration and confine the collected trajectories to rapidly failing episodes.
Excluding Herald from the data collection instead enables it to acquire the basic reconstruction ability quickly. 
Moreover, it supervises Herald on the trajectories that the MAS is supposed to follow, which are exactly the ones Herald is ultimately aiming to restore.

\emph{Stage 2: On-Policy Distillation.}
Stage 1 alone is not sufficient, since incorporating Herald will change the trajectories that the MAS follows, making the distribution of the target messages drift.
To account for the drifted distribution, we leverage the idea of Dataset Aggregation~(DAgger)~\cite{ross2011dagger} and train Herald on the trajectories that itself induces.
Specifically, the trained Herald in Stage 1 is inserted into the MAS's execution loop, and trajectories are collected under packet loss to form dataset $\mathcal{D}_{\mathrm{on}}$ in this on-policy condition. 
Moreover, to prevent the reconstruction ability learned in Stage 1 from being forgotten, we replay part of the Stage-1 experience, and thus the training set is given by
\begin{align}
\mathcal{D}_{\mathrm{mix}}
=
\left(1-\rho\right)\mathcal{D}_{\mathrm{on}}
+
\rho\,\mathcal{D}_{\mathrm{off}},
\label{eq:Herald_mixed_dataset}
\end{align}
where $\rho\in[0,1]$ is the mixing ratio.

\subsubsection{Training the Verification Capability}
\label{subsec:Herald_verifier_training}

With the prediction capability established, we train the verifier while keeping the MAS parameters and the trained prediction-related modules frozen.
Again, we first formulate the objective and then describe the training method.

\textbf{Objective Function Formulation.}
During training, the simulator selects received teammate packets with a prescribed probability and replaces their contents with adversarial messages.
The generation of adversarial messages is described in Sec.~\ref{subsec:setup_Herald}.
The verifier assesses each received teammate message following the procedures described in Sec.~\ref{subsec:Herald_verifier}.
To optimize the verifier efficiently, we augment the classification loss with two additional terms for learning accurate individual assessments and reference predictions, yielding
\begin{align}
    \mathcal L_{\mathrm{ver}}
    &=\mathcal L_{\mathrm{cls},\mathrm{fin}}
    +\beta_{\mathrm{ind}}\mathcal L_{\mathrm{cls},\mathrm{ind}}
    +\beta_{\mathrm{ref}}\mathcal L_{\mathrm{ref}},
    \label{eq:Herald_verifier_training_loss}
\end{align}
where $\beta_{\mathrm{ind}}$ and $\beta_{\mathrm{ref}}$ are non-negative weights to balance multiple objectives.
Here, $\mathcal L_{\mathrm{cls},\mathrm{fin}}$ and $\mathcal L_{\mathrm{cls},\mathrm{ind}}$ are the mean binary cross-entropy losses of final and individual trustworthiness scores in~\eqref{eq:Herald_verifier_final_score} and~\eqref{eq:Herald_verifier_initial_score_1} against the binary ground-truth authenticity label.
The reference loss $\mathcal L_{\mathrm{ref}}$ compares the predicted reference in~\eqref{eq:Herald_verifier_prior} with the original soft distributions recorded before manipulation.
To preserve reference prediction as the learning objective of the reference constructor, we stop classification gradients at the four statistics in~\eqref{eq:Herald_verifier_statistics}, so that the constructor's parameters are updated only by $\mathcal L_{\mathrm{ref}}$.

\textbf{Training Method.}
Verifier training follows the two-stage distillation procedure used for the prediction capability with the adaptations detailed below.\looseness=-1\relax

\emph{Adaptation in Stage 1.}
We reuse the lossless trajectories in $\mathcal D_{\mathrm{off}}$ while introducing synthetic message manipulation alongside packet loss during replay, and optimize the verifier using~\eqref{eq:Herald_verifier_training_loss}.
The original messages provide authenticity labels for the received candidates, while their recorded soft distributions serve as distillation targets for the reference constructor.\looseness=-1\relax

\emph{Adaptation in Stage 2.}
We collect trajectories with the verifier trained in Stage-1 being activated to screen messages at the preset threshold $\tau_{\mathrm{ver}}$.
We then mix these trajectories with $\mathcal D_{\mathrm{off}}$ according to~\eqref{eq:Herald_mixed_dataset} and continue optimizing the verifier for~\eqref{eq:Herald_verifier_training_loss}, with the same supervision as in Stage 1.

In summary, the complete training algorithm for Herald is provided in Algorithm~\ref{alg:Herald_training}.

\begin{algorithm}[t]
\setlength{\algomargin}{0pt}
\rightskip=0pt plus 0.3em\relax
\caption{Two-stage Distillation-based Training Algorithm for Herald.}
\label{alg:Herald_training}
\KwInput{Frozen MAS parameters $\Phi$, packet-loss rate $p$, data mixing ratio $\rho$, and adversarial message rate $p_{\mathrm{A}}$, acceptance threshold $\tau_{\mathrm{ver}}$.}
\KwOutput{Trained parameters of message encoder ($\theta_{\mathrm{enc}}$), tracker ($\theta_{\mathrm{tra}}$), predictor ($\theta_{\mathrm{pre}}$), and verifier ($\theta_{\mathrm{verf}}$).}
\textbf{\# Training the prediction capability}\;
\emph{Stage 1:} Collect lossless trajectories $\mathcal D_{\mathrm{off}}$ with original messages and soft targets, and train $\theta_{\mathrm{enc}}$, $\theta_{\mathrm{tra}}$, and $\theta_{\mathrm{pre}}$ on $\mathcal D_{\mathrm{off}}$ to minimize~\eqref{eq:Herald_total_loss} under simulated packet loss at rate $p$\;
\emph{Stage 2:} Collect $\mathcal D_{\mathrm{on}}$ with the trained prediction components under packet loss, and continue training for~\eqref{eq:Herald_total_loss} on $\mathcal D_{\mathrm{mix}}$ defined in~\eqref{eq:Herald_mixed_dataset}\;
\textbf{\# Training the verification capability}\;
\emph{Stage 1:} Freeze $\theta_{\mathrm{enc}}$, $\theta_{\mathrm{tra}}$, and $\theta_{\mathrm{pre}}$, and train $\theta_{\mathrm{verf}}$ on $\mathcal D_{\mathrm{off}}$ to minimize~\eqref{eq:Herald_verifier_training_loss} under simulated packet loss at rate $p$ and message manipulation at rate $p_{\mathrm A}$\;
\emph{Stage 2:} Collect new $\mathcal D_{\mathrm{on}}$ under the same conditions with verifier screening at $\tau_{\mathrm{ver}}$, and continue training $\theta_{\mathrm{verf}}$ for~\eqref{eq:Herald_verifier_training_loss} on the resulting $\mathcal D_{\mathrm{mix}}$\;
\end{algorithm}

%% file: sections/5_experiment_setup.tex
\section{Experiment Setup}
\label{sec:setup}

In this section, we describe the setup under which Herald is evaluated, covering the benchmark environment, the MAS that Herald is attached to, the configuration used to train Herald and its verifier, and the protocol under which all reported results are measured.

\subsection{StarCraft Multi-Agent Challenge}
\label{subsec:setup_smac}
We evaluate Herald on the SMAC~\cite{samvelyan2019smac}, a standard benchmark for decentralized micromanagement in StarCraft II, in which one agent controls one allied unit and the resulting team fights an enemy army commanded by the built-in game AI at its highest non-cheating difficulty. 
As the units move continuously and perceive only their surroundings, SMAC serves as an abstraction of a team of mobile embodied agents. 
Each agent perceives only the units inside its circular sight range, together with their relative positions, health, shields, and types, and is blind to everything beyond it, which is exactly what makes the messages in \eqref{eq:Herald_message} necessary. 
The discrete action space comprises no-op, stop, movement in four cardinal directions, and one attack action per enemy unit, and is hard-masked at every step to the currently available actions, so that only legal actions are executed. All agents share a single scalar reward accumulated from the damage dealt, the units eliminated, and a bonus for defeating the other team. We evaluate Herald on five scenarios: \texttt{3s\_vs\_3z}, \texttt{8m}, \texttt{2s3z}, \texttt{5m\_vs\_6m}, and \texttt{corridor}.

\subsection{Base MAS Implementation}
\label{subsec:setup_base}
We instantiate the underlying MAS with the framework of~\cite{Egorov2022MAMBA}, following the execution pipeline of Sec.~\ref{subsec:wm_comm_marl}. The latent state consists of $32$ categorical variables, each with $32$ categories, whereas the deterministic state, the observation embedding, and the hidden layers all use a width of $256$; the remaining hyperparameters follow the SMAC configuration of~\cite{Egorov2022MAMBA}. For every scenario, a separate base model is trained from scratch under ideal communication, i.e., every message is delivered at every step. Training runs for $100$k environment steps on \texttt{3s\_vs\_3z}, \texttt{8m}, and \texttt{2s3z}, $200$k steps on \texttt{5m\_vs\_6m}, and $500$k steps on \texttt{corridor}. The resulting world model, actor, and critic are then frozen, so that Herald is trained and deployed on top of them without altering a single parameter.\looseness=-1\relax

\subsection{Training Configuration}
\label{subsec:setup_Herald}
\textbf{Prediction Training.}
Herald uses a message encoder and a readout of width $256$, and its latent state tracker consists of two Mamba-3 blocks~\cite{lahoti2026mamba3}.
In Stage~1, the frozen controller is rolled out under ideal communication to collect an off-policy dataset $\mathcal{D}_{\mathrm{off}}$ of $240$ episodes, on which packet losses are synthesized offline at the target rate $p$, and \eqref{eq:Herald_total_loss} is optimized for $3000$ gradient steps, with $\beta=0.1$.
In Stage~2, the resulting module is placed in the execution loop over the same channel to collect an on-policy dataset $\mathcal{D}_{\mathrm{on}}$ of another $240$ episodes, whose lossless messages are retained as supervision targets.
Herald is then updated for $800$ steps on the mixture $\mathcal{D}_{\mathrm{mix}}$ of \eqref{eq:Herald_mixed_dataset}, drawing six on-policy and two off-policy subsequences per batch, i.e., $\rho = 0.25$.

\textbf{Verification Training.}
When training the verifier, additional manipulated messages are randomly injected into the team communication to replace the packets that should have been received, and each injected message is assumed to always arrive.
For control of variables, we keep the statistics of authentic packet arrivals unchanged in the adversarial environment, so that each directed teammate link delivers the authentic message, a manipulated one, or nothing with probabilities $1-p$, $p_{\mathrm A}$, and $p-p_{\mathrm A}$, respectively.
We consider three typical types of adversarial manipulations.
\emph{Flip} selects $B_{\mathrm{flip}}\in[10,30]$ fields from the $32+1$ variables in a message and changes each to a legal category different from its original one.
\emph{Replay} substitutes the sender's own authentic message from $B_{\mathrm{repl}}\in[10,30]$ communication steps earlier, delivered as if it were current.
\emph{Grad} replaces up to $B_{\mathrm{grad}}\in[10,30]$ fields with the categories of largest positive first-order effect on the policy's deviation from its original action.
We note that \emph{Grad} requires the adversary to know the received messages of the receiver and the internal state of its tracker, thereby constituting an idealized white-box attack.\looseness=-1\relax

We train a dedicated verifier for each manipulation on top of the same prediction module, which is trained by the prediction training and kept frozen.
Stage~1 optimizes \eqref{eq:Herald_verifier_training_loss} for $3000$ updates on $\mathcal{D}_{\mathrm{off}}$, with losses and attacks synthesized offline at a per-trajectory rate $p_{\mathrm A}$ uniformly sampled in $[0.1,0.5]$.
Stage~2 deploys the Stage-1 endpoint at $\tau_{\mathrm{ver}}=0.5$ to collect $240$ fresh episodes and continues for $800$ updates on the $\rho=0.25$ mixture of \eqref{eq:Herald_mixed_dataset}.

\subsection{Evaluation Protocol}
\label{subsec:setup_protocol}
All methods are evaluated on the same $1000$ episodes, whose seeds are mutually distinct and disjoint from every seed used for base training and for Herald data collection.
Episodes are capped at $400$ steps.
Within each episode, the seeds for the channel, the latent-state sampling, and the adversarial manipulations (when present) are fixed, so that runs of the same episode under different methods are directly comparable.
By default, each directed teammate link drops packets i.i.d. with $p=0.9$, while the self-link is always available.

We report the win rate together with its Wilson score confidence interval (CI), which for $v$ victories out of $n$ episodes with $\hat{w}=v/n$ reads
\begin{align}
\mathrm{CI} = \frac{1}{1+c_{\alpha}^{2}/n}
\big(
\hat{w}+\frac{c_{\alpha}^{2}}{2n}
\pm
c_{\alpha}\sqrt{\frac{\hat{w}\left(1-\hat{w}\right)}{n}+\frac{c_{\alpha}^{2}}{4n^{2}}}
\big),
\label{eq:wilson}
\end{align}
where $c_{\alpha}$ is the standard normal quantile at level $1-\alpha/2$.
We set $n=1000$ and $\alpha=0.05$, so that $c_{\alpha}\approx1.96$, and refer to the resulting interval as the \emph{Wilson $95\%$ CI}, which characterizes the episode-level sampling uncertainty of a fixed checkpoint at $95\%$ confidence.
We report Wilson $95\%$ CI as error bars or shaded regions in the figures in Sec.~\ref{sec:evaluation}.

%% file: sections/6_evaluation.tex
\section{Evaluation Results}
\label{sec:evaluation}

In this section, we evaluate the performance of Herald.
We first compare Herald with existing packet-loss handling methods across scenarios, then ablate its components and alternative tracker architectures, analyze the factors that affect its performance, namely the tracker hyperparameters and the deployment packet-loss process, and finally examine its robustness to adversarial message manipulations.

\input{Figures/experiment/tab_overall_compare_0728}

\textbf{Overall Comparison.} Table~\ref{tab:overall_compare} compares Herald with six packet-loss baselines across five scenarios, spanning symmetric and asymmetric engagements, MAS sizes from three to eight agents, and both homogeneous and heterogeneous agent roles.
The first column lists the SMAC scenarios, whose details can be found in~\cite{samvelyan2019smac}, and the second column summarizes their attributes. 
\texttt{3s\_vs\_3z} is labeled weak because each agent can individually kite one adversarial entity to win, whereas the remaining scenarios rely more on team coordination. 

The third to tenth columns report the win rates achieved by different methods.
\emph{Ideal} denotes the lossless communication under which the MAS was trained, and therefore upper-bounds the attainable performance.
\emph{MARO}~\cite{Santos2025MARO} and \emph{MATCH}~\cite{Ding2026MATCH} are two recent state-of-the-art methods for operating a MAS under non-ideal communication.
Both share the core idea of predicting the lost packet from history and available teammate packets that survived, but differ in predictor architecture and training objective. 
MARO uses an LSTM predictor that regresses the incremental team state under a Gaussian predictive likelihood loss, whereas MATCH uses a Transformer predictor that recursively interpolates the missing packet while optimizing message reconstruction accuracy together with the value prediction accuracy.
Both assume the MAS to be re-trained under the non-ideal channel from scratch; however, for a fair comparison, we retain their core mechanisms but train only their communication-fault-tolerance modules on top of the frozen MAS.\!\footnote{We also tried jointly fine-tuning the agents together with these modules for $10$k steps, but both variants performed far worse, with win rates $\leq 1\%$.}
The remaining four columns are typical training-free methods: 
\emph{ZOH}~(zero-order holding) holds the most recently received packet of each sender.
\emph{Roll-out} leverages the world model to predict the lost latent states from their prior.
\emph{Mask-out} makes the communication module of the MAS omit the lost packets.
Finally, \emph{No-comm} indicates no communication, where all teammate packets are discarded.

Herald achieves the highest win rate among all methods operating under $p=0.9$ packet loss in every scenario. Its gap to the lossless Ideal is only $2.6$, $1.5$, and $1.4$ percentage points on \texttt{8m}, \texttt{2s3z}, and \texttt{3s\_vs\_3z}, respectively, and remains $9.2$ and $5.4$ points on the more difficult \texttt{5m\_vs\_6m} and \texttt{corridor}.
Averaged across the five scenarios, Herald multiplies the Mask-out win rate by $2.16$, clearly ahead of the second-best method MATCH at $1.61$, and close to the $2.46$ attained by the lossless Ideal; MARO reaches $1.21$, while ZOH and Roll-out fall below one at $0.84$ and $0.46$, respectively.
Compared with the Ideal, Herald recovers $92.1\%$ of the Ideal win rate, whereas MATCH, MARO, and Mask-out recover only $76.7\%$, $64.2\%$, and $60.1\%$.
Overall, Herald substantially outperforms all compared methods and its margin widens precisely on the scenarios that demand the strongest coordination.

\begin{figure}[t]
    \centering
    \includegraphics[width=\columnwidth]{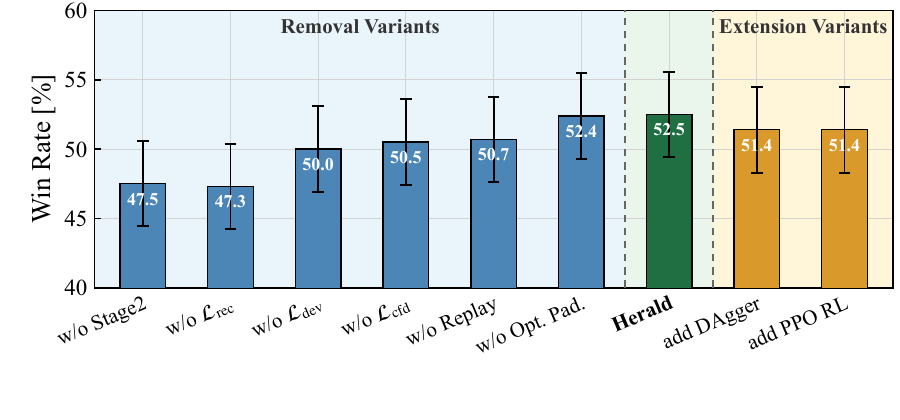}
    \vspace{-3.5em}
    \caption{Ablation of Herald on \texttt{corridor} under $p=0.9$ by removing its core components or extending it with extra training stages.}
    \label{fig:herald_ablation}
    \vspace{-1em}
\end{figure}

\textbf{Algorithm Ablation.} Fig.~\ref{fig:herald_ablation} shows the ablation results of the components of Herald evaluated on \texttt{corridor}.
The \emph{removal variants} on the left each remove one component, namely the Stage~2, the reconstruction loss $\mathcal{L}_{\mathrm{rec}}$, the deviation loss $\mathcal{L}_{\mathrm{dev}}$, the confidence loss $\mathcal{L}_{\mathrm{cfd}}$, the experience replay, and the optimized padding for lost messages.
Every removal degrades the win rate, which shows that each component makes a positive contribution to Herald.
The Stage~2 training and $\mathcal{L}_{\mathrm{rec}}$ have the largest impact, contributing $5.0\%$ and $5.2\%$ of win rate, respectively, whereas the optimized padding contributes only $0.1\%$, with the remaining components lying in between.
The extension variants on the right instead augment Herald with one additional round of DAgger training or with an equal number of PPO~\cite{schulman2017ppo} optimization steps applied to the predictor and the tracker.
However, they slightly degrade the win rate, which we attribute to the instability caused by extra training.
These results justify the design of Herald.

\begin{figure}[t]
    \centering
    \includegraphics[width=\columnwidth]{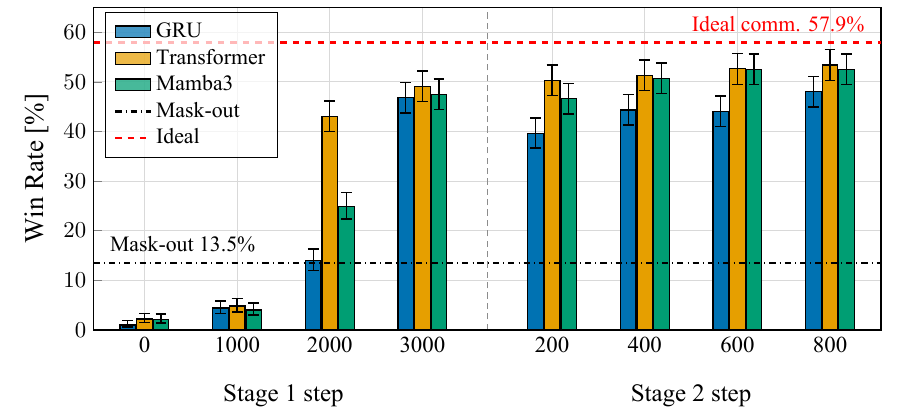}
    \vspace{-2.em}
    \caption{Comparison of tracker architectures on \texttt{corridor} under $p=0.9$.}
    \label{fig:archi_compare}
    \vspace{-1em}
\end{figure}

\textbf{Neural Architecture Comparison.} Fig.~\ref{fig:archi_compare} reports the win rate of Herald over the two training stages when its tracker is implemented by different neural network architectures.
We compare our selected Mamba-3 against the classical GRU~\cite{cho2014gru} and Transformer~\cite{vaswani2017transformer}, with the three configured to have matched parameter counts of about 3 million.
The Transformer maintains an explicit context window over the most recent $128$ steps.
This gives it the fastest Stage~1 learning speed and the highest final win rate of $53.4\%$, but it is also the most expensive: at a full window its single-step cost reaches $2.42$~GFLOPs, about $19.7\times$ that of the Mamba-3 and about $21.8\times$ that of GRU.
As architectures that instead evolve a fixed-size recurrent state, Mamba-3 and GRU remain far cheaper in terms of computational complexity.
Between the two, Mamba-3 achieves higher performance than GRU, as its complex-valued state transitions and exponential-trapezoidal recurrence allow more precise modeling of the latent dynamics.
Among the three architectures, the default-adopted Mamba-3 offers the best trade-off between computational complexity and achievable performance.\looseness=-1\relax

\begin{figure}[t]
    \vspace{-2.5ex}
    \centering
    \subfloat[]{%
        \includegraphics[width=0.5\columnwidth]{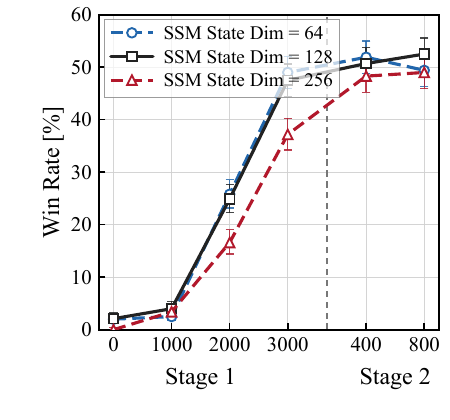}}
    \subfloat[]{%
        \includegraphics[width=0.5\columnwidth]{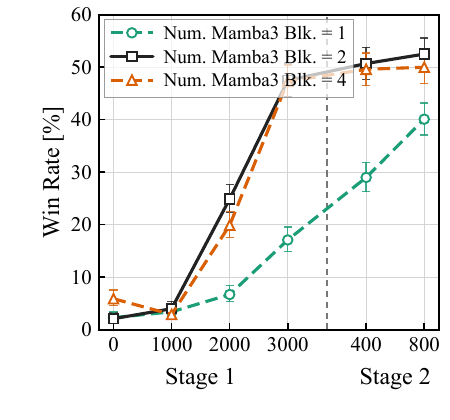}}
    \caption{Impact of (a) SSM state dimension and (b) number of Mamba-3 blocks on Herald in \texttt{corridor}, under $p=0.9$.}
    \label{fig:hyperparam_impact}
    \vspace{-1em}
\end{figure}

\textbf{Impact of Tracker Capacity.} Fig.~\ref{fig:hyperparam_impact} shows how the capacity of the Mamba-3 tracker affects the win rate, tested on \texttt{corridor} under $p=0.9$.
We focus on two hyperparameters that directly determine this capacity.
The first is the SSM state dimension, which sets the width of the recurrent memory matrix and thus how much information each receiver can retain.
The second is the number of stacked Mamba-3 blocks introduced in Sec.~\ref{subsec:Herald_architecture}, which controls the depth of the modeled latent dynamics.
Herald adopts a state dimension of $128$ and two blocks by default, amounting to $1.37$M tracker parameters.
A smaller state dimension leads to faster training, whereas too few blocks slow the improvement down instead, as they are less capable of expressing the latent dynamics.
Meanwhile, a larger capacity, in either state dimension or number of blocks, brings no significant gain, which we attribute to the stochasticity inherent in neural network training and to the greater optimization difficulty introduced by more parameters.

\begin{figure}[t]
    \centering
    \subfloat[]{%
        \label{fig:pkt_loss_impact-a}%
        \includegraphics[width=0.49\columnwidth]{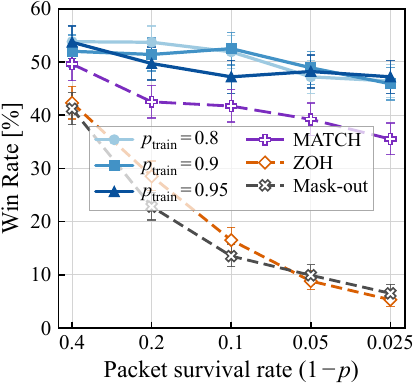}}
    \hfil
    \subfloat[]{%
        \label{fig:pkt_loss_impact-b}%
        \includegraphics[width=0.49\columnwidth]{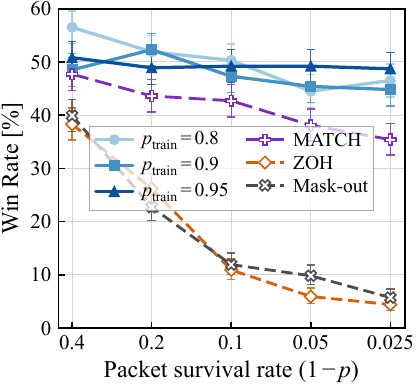}}
    \caption{Impact of the deployment packet-loss rate on Herald in \texttt{corridor} under (a) i.i.d. Bernoulli and (b) GE packet loss. MATCH is trained under $p=0.9$.} \label{fig:pkt_loss_impact}
    \vspace{-1em}
\end{figure}

\textbf{Impact of Deployment Channels.}
Fig.~\ref{fig:pkt_loss_impact-a} examines how Herald behaves when the packet-loss rate encountered at deployment differs from the one it was trained under.
Specifically, we take Heralds trained at $p=0.8$, $0.9$, and $0.95$, and deploy each of them over Bernoulli packet channels whose actual loss rate spans $0.6$ to $0.975$, equivalent to a packet survival rate spanning $0.4$ to $0.025$.
As the number of evaluated episodes is limited and the training of neural networks suffers from randomness, the differences among the three Herald curves are not pronounced; nevertheless, two observations can be drawn.
First, compared with ZOH and Mask-out, every Herald curve decreases much more slowly as the loss rate increases. 
Herald also consistently outperforms MATCH, the second-best in Table~\ref{tab:overall_compare}, by a notable margin.
Second, a Herald trained at one loss rate generalizes to both lower and higher loss rates and can achieve comparable performance with the Herald trained under those conditions.

To further test this generalizability of Herald, in Fig.~\ref{fig:pkt_loss_impact-b}, we deploy the same three trained Heralds under Gilbert--Elliott (GE) channels~\cite{mushkin1989gilbert}, where the packet losses are bursty instead of independent, as is typical when mobile agents are temporarily blocked or shadowed.
We set the correlation coefficient of the GE channel to be $0.3$ while keeping the same average loss rates used in Fig.~\ref{fig:pkt_loss_impact-a}, increasing the expected gap length by a factor of $1.43$.
As shown, Herald trained under i.i.d. Bernoulli loss also transfers to the bursty GE channel, demonstrating that the tracker learns latent dynamics rather than the statistics of one specific loss process.

\begin{figure}[t]
    \vspace{-2.5ex}
    \centering
    \subfloat[]{%
        \includegraphics[width=0.46\columnwidth]{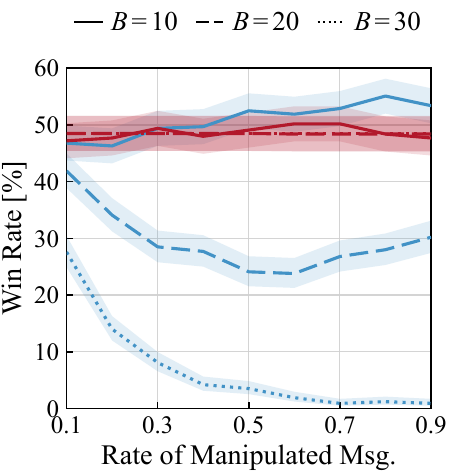}}
    \hfil
    \subfloat[]{%
        \includegraphics[width=0.46\columnwidth]{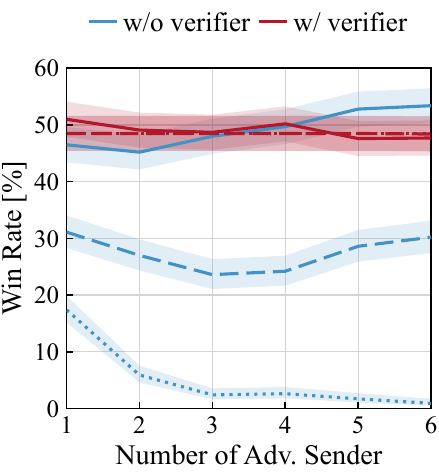}}\\
    \subfloat[]{%
        \includegraphics[width=0.46\columnwidth]{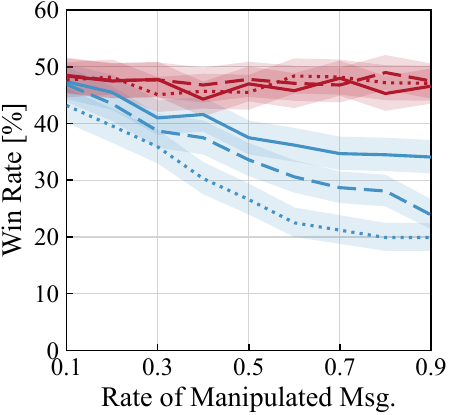}}
    \hfil
    \subfloat[]{%
        \includegraphics[width=0.46\columnwidth]{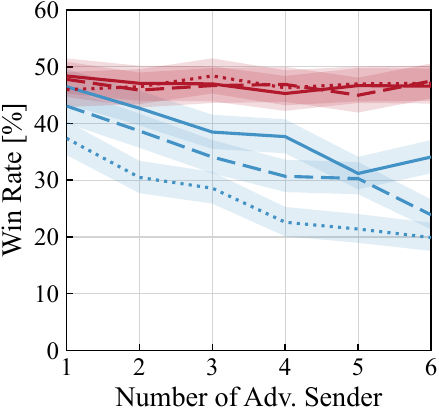}}\\
    \subfloat[]{%
        \includegraphics[width=0.46\columnwidth]{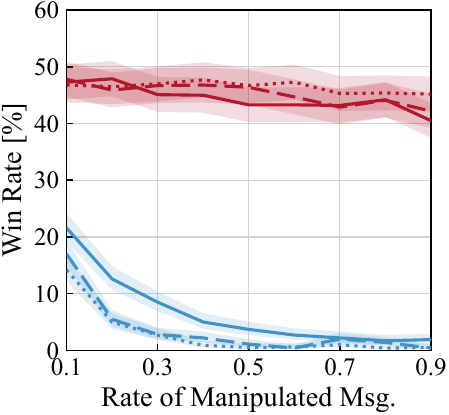}}
    \hfil
    \subfloat[]{%
        \includegraphics[width=0.46\columnwidth]{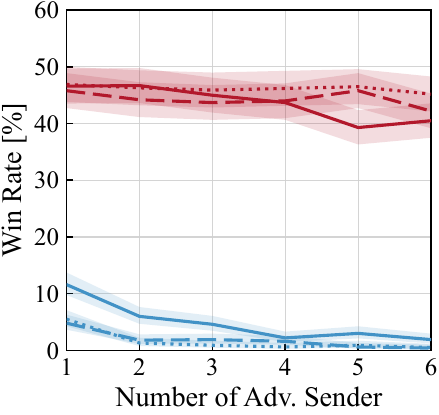}}
    \caption{Comparison of the win rates of Herald with and without its dedicated verifier under \emph{Flip} (a) and (b), \emph{Replay} (c) and (d), and \emph{Grad} adversarial attacks (e) and (f), tested on \texttt{corridor} at packet loss rate $p=0.9$.}
    \label{fig:injection_robustness}
    \vspace{-1em}
\end{figure}

\textbf{Robustness to Adversarial Manipulations.}
Finally, we evaluate Herald and its verifier in an adversarial environment, with injected manipulated messages of the three types defined in Sec.~\ref{subsec:setup_Herald}.
The evaluation is conducted on the MAS for \texttt{corridor} under $p=0.9$.
For each type of manipulation, its dedicated verifier is employed to screen messages.
The attack budgets of the three manipulations, i.e., $B_{\mathrm{flip}}$, $B_{\mathrm{repl}}$, and $B_{\mathrm{grad}}$ (uniformly denoted by $B$), take values in $\{10,20,30\}$.
We consider a link-level and an agent-level scenario.
In the link-level scenario, the communication between any pair of agents receives manipulated messages at a rate $p_{\mathrm A}$ varying from $0.1$ to $0.9$.
In the agent-level scenario, $m\in\{1,\dots,6\}$ of the six agents are randomly hijacked by an adversary, and the messages arriving at their receivers are manipulated with probability $0.9$.\footnote{We assume a strong adversary whose injected manipulated messages are not affected by the environmental packet-loss process.}
We note that only the link-level scenario with $p_{\mathrm A}\in[0.1,0.5]$ falls within the training conditions of the verifier, whereas the remaining link-level conditions and the whole agent-level scenario lie outside the training distribution.
By this means, we test the generalization of the verifier.

Fig.~\ref{fig:injection_robustness} reports the win rates of Herald, with and without the verifier, tested in the two scenarios.
All sub-figures share the same no-attack baseline, namely Herald's average win rate of $52.5\%$ on \texttt{corridor} in Table~\ref{tab:overall_compare}.
Overall, the performance of Herald without the verifier shows that \emph{Grad} is the most effective attack: even at the lowest budget $B_{\mathrm{grad}}=10$ and the lowest rate $p_{\mathrm A}=0.1$, it significantly degrades the performance of the MAS.
\emph{Flip} is the second most effective; however, at $B_{\mathrm{flip}}=10$, the win rate of Herald without the verifier can exceed that with the verifier and even the no-attack baseline.
This is because, at this budget, each manipulated message differs from the authentic one in only $10$ of its $33$ fields and thus can provide the MAS with much additional accurate information.
The same reasoning explains the performance recovery of $B_{\mathrm{flip}}=20$ at high injection rates.
\emph{Replay} is the least effective attack, since the replayed messages, though outdated, remain authentic in content and thus informative.
Notably, equipping the corresponding verifier restores the MAS's performance in adversarial environments.
The restoration persists under the manipulated-message rates beyond the training range and under the agent-level adversary which is out of the training distribution.
The results validate the design of the verifier as well as the generalizability that it obtains from limited training.\looseness=-1\relax

%% file: Figures/experiment/tab_overall_compare_0728.tex
\begin{table*}[!t]
    \centering
    \caption{Win rates (\%) comparison on five SMAC scenarios at $p=0.9$}
    \label{tab:overall_compare}
    \begingroup
    \footnotesize
    \setlength{\tabcolsep}{2.0pt}
    \setlength{\doublerulesep}{1.2pt}
    \renewcommand{\arraystretch}{1.3}
    \begin{tabular}{@{}l l||c c c c c c c c@{}}
        \hline\hline
        \textbf{Scenario} &
        \textbf{Attributes} &
        {Ideal} &
        \textbf{Herald} &
        {MARO} &
        {MATCH} &
        {ZOH} &
        {Roll-out} &
        {Mask-out} &
        {No-comm} \\
        \hline
        \texttt{8m} &
        8 agents, homogeneous, strong cooperation &
        $81.8_{-2.5}^{+2.3}$ &
        $\bm{79.2_{-2.6}^{+2.4}}$ &
        $76.0_{-2.7}^{+2.5}$ &
        $75.8_{-2.8}^{+2.6}$ &
        $76.3_{-2.7}^{+2.5}$ &
        $36.8_{-2.9}^{+3.0}$ &
        $74.3_{-2.8}^{+2.6}$ &
        $51.6_{-3.1}^{+3.1}$ \\
        \texttt{2s3z} &
        5 agents, heterogeneous, strong cooperation &
        $80.8_{-2.6}^{+2.3}$ &
        $\bm{79.3_{-2.6}^{+2.4}}$ &
        $51.2_{-3.1}^{+3.1}$ &
        $73.6_{-2.8}^{+2.6}$ &
        $61.4_{-3.1}^{+3.0}$ &
        $36.0_{-2.9}^{+3.0}$ &
        $61.9_{-3.1}^{+3.0}$ &
        $50.6_{-3.1}^{+3.1}$ \\
        \texttt{3s\_vs\_3z} &
        3 agents, homogeneous, weak cooperation &
        $93.0_{-1.8}^{+1.4}$ &
        $\bm{91.6_{-1.9}^{+1.6}}$ &
        $84.5_{-2.4}^{+2.1}$ &
        $82.6_{-2.5}^{+2.2}$ &
        $74.0_{-2.8}^{+2.6}$ &
        $51.6_{-3.1}^{+3.1}$ &
        $81.4_{-2.5}^{+2.3}$ &
        $70.9_{-2.9}^{+2.7}$ \\
        \texttt{5m\_vs\_6m} &
        5 agents, homogeneous, strong cooperation &
        $39.3_{-3.0}^{+3.1}$ &
        $\bm{30.1_{-2.8}^{+2.9}}$ &
        $0.8_{-0.4}^{+0.8}$ &
        $15.2_{-2.1}^{+2.4}$ &
        $0.5_{-0.3}^{+0.7}$ &
        $0.0_{+0.0}^{+0.4}$ &
        $8.8_{-1.6}^{+1.9}$ &
        $2.5_{-0.8}^{+1.2}$ \\
        \texttt{corridor} &
        6 agents, homogeneous, strong cooperation &
        $57.9_{-3.1}^{+3.0}$ &
        $\bm{52.5_{-3.1}^{+3.1}}$ &
        $41.6_{-3.0}^{+3.1}$ &
        $41.7_{-3.0}^{+3.1}$ &
        $16.5_{-2.2}^{+2.4}$ &
        $8.1_{-1.5}^{+1.9}$ &
        $13.5_{-2.0}^{+2.3}$ &
        $6.1_{-1.3}^{+1.7}$ \\
        \hline\hline
    \end{tabular}
    \endgroup\\[1ex]
    * Subscripts and superscripts are the lower and upper Wilson $95\%$ CI deviations over $1000$ episodes, as defined in \eqref{eq:wilson}.
\vspace{-1em}
\end{table*}

%% file: sections/7_conclusion.tex
\section{Conclusion}
\label{sec:conclusion}
This paper addressed how to retrofit a pretrained mobile communicative MAS with resilience to extreme packet loss and adversarial message manipulation without retraining its underlying policy.
We proposed Herald, a lightweight receiver-side plugin that combines a Mamba-3-based tracker with a predictor to reconstruct missing messages and estimate their confidence, and a verifier that screens adversarially manipulated messages before they distort the tracker's history.
We then designed the two-stage distillation method that combines off-policy supervision on lossless trajectories with on-policy refinement to mitigate deployment-time covariate shift.
Experiments on SMAC show that Herald consistently outperforms multiple benchmarks and baselines, recovers an average of 92.1\% of the ideal-communication win rate under 90\% packet loss, generalizes to unseen loss rates and bursty channels, and retains performance under three types of adversarial attacks even beyond the training distribution.\looseness=-1\relax